\documentclass[reprint,amsmath,amssymb,aps,prb]{revtex4-1}
\usepackage{graphicx}
\usepackage[utf8]{inputenc}
\usepackage[T1]{fontenc}
\usepackage{mwe}
\usepackage{epsfig}
\usepackage{graphicx}
\usepackage{subcaption}
\usepackage{adjustbox} 
\usepackage{verbatim}
\usepackage{bm}
\usepackage{cleveref}
\usepackage[compact]{titlesec}

\begin{document}
\title{Characterization of Lattice Thermal Transport in Two-Dimensional BAs, BP, and BSb: A First-Principles Study}
\author{Charles Shi, Xuan Luo*}
\affiliation{National Graphene Research and Development Center, Springfield, Virginia 22151, USA}
\date{\today}
\begin{abstract}
\setlength{\parindent}{1cm}
Effective thermal management and high thermal conductivity mechanisms are becoming more and more crucial in today's more demanding electronic circuits. Ever since the high thermal conductivity in cubic boron arsenide (c-BAs) was predicted theoretically by Lindsay et. al in 2013, countless studies have zeroed in on this particular material. Most recently, c-BAs has been confirmed experimentally to have a thermal conductivity of around 1,100 W/m-K. In this study, we investigate the seldom studied two dimensional hexagonal form of boron arsenide (h-BAs) using a first-principles approach and by solving the Boltzmann Transport Equation for phonons. Traditionally, a good indicator of a high thermal conductivity material is its high Debye temperature and high phonon group velocity. However, we determine h-BAs to have a much lower Debye temperature and average phonon group velocity compared to the other monolayer boron-V compounds of boron nitride (h-BN) and boron phosphide (h-BP), yet curiously it possesses a higher thermal conductivity. Further investigation reveals that this is due to the phonon frequency gap caused by large mass imbalances, which results in a restricted Umklapp phonon-phonon scattering channel and consequently a higher thermal conductivity. We determine the intrinsic lattice thermal conductivity of monolayer h-BAs to be 362.62 W/m-K at room temperature, which is considerably higher compared to the other monolayer boron-V compounds of h-BN (231.96 W/m-K), h-BP (187.11 W/m-K), and h-BSb (87.15 W/m-K). This study opens the door for investigation into a new class of monolayer structures and the properties they possess.

\end{abstract}

\newpage 
\maketitle
\bigskip
\section{INTRODUCTION}
\bigskip
	As the demand for electric power increases in today's more advanced electronics, the demand for electronics cooling and therefore materials with high thermal conductivity, $\kappa$, increases as well. Two-dimensional materials are gaining momentum for these applications because of their capability to be implemented in a wide range of next-generation electronic devices.\cite{sivacarendran2014, balandin2008, lee2011, dai2016, balatero2015, peng2016low, peng2016phonon, qin2015, li2017, shafique2017ultra, peng2016towards, xiao2017, liu2017, ma2016, gonzalez2018, khatami2016, jiang2017, shafique2017thermoelectric, kuang2016, wei2011, ong2014, pei2013, chen2012, guo2016, jiang2011, das2017, li2012, gu2015} Graphene, a single layer of carbon atoms arranged in a honeycomb lattice, has been confirmed to have an ultrahigh $\kappa$ that is able to outperform bulk thermal conductors like its expensive and rare cousins of graphite and diamond \cite{balandin2008, seol2010}. Despite its high $\kappa$, however, graphene's lack of electronic bandgap makes it highly suboptimal for ideal implementation in many electronic devices.\cite{sivacarendran2014, wang2017, dai2016} In this study, every material we investigate has a direct electronic bandgap, thus having more realistic applications with today's electronics than graphene.\cite{sahin2009}

	Recently, the experimental observations of ultrahigh $\kappa$ in bulk, cubic boron phosphide (c-BP)\cite{kang2017} and boron arsenide (c-BAs)\cite{kang2018, li2018, tian2018} have ignited new flames in the discovery of high $\kappa$ materials. c-BP and c-BAs were shown to have experimental values of approximately 400 and 1,100 W/m-K at room temperature, respectively. These values, although smaller in magnitude than the original theoretical prediction by Boltzmann theory,\cite{broido2013, lindsay2013} are still considered very high -- regarded here as exceeding that of copper, or 400 W/m-K at room temperature.\cite{dames2018}  This study simulates the bulk form as a two dimensional, hexagonal lattice of differing boron and group-V atoms, and our results observe the $\kappa$ of monolayer h-BP and h-BAs to be 187.11 and 362.26 W/m-K at room temperature, respectively. We observe the thermal conductivity of h-BAs to be only slightly lower than this 400 W/m-K threshold, which could be as a result of the fundamental shortcomings in the approach we follow which we will examine further on.

	In 1973, Slack pinpointed four properties of high thermal conductivity non-metallic crystals, where thermal transport is dominated by phonons: (I) the material must have a low average atomic mass, (II) the material must have strong interatomic bonding, (III) the material must have a simple crystal structure, and (IV) the material must have low anharmonicity.\cite{slack1973} These four properties, which often conflict with one another,\cite{peng2016thermal} govern the chances of high thermal conductivity material. Conditions I and II are dictated by a high Debye temperature, condition III implies a small number of atoms per unit cell, and condition IV is quantified by a low Gr\"uneisen parameter. c-BAs has been shown to be a rather large anomaly when considering this theory, as it has an exceptionally high thermal conductivity, despite its heavy average atomic mass (42.87 amu) and its low Debye temperature (700 K) when compared to other high $\kappa$ materials.\cite{lindsay2013} Resembling its bulk form, monolayer h-BAs follows the same trend, as its low Debye temperature (435.62 K) also contrasts with higher $\kappa$ when compared to other monolayer materials. A known explanation for this anomaly in the bulk form is the large acoustic-optical phonon band gap, which is also reflected in the monolayer form.

	We follow a well-known approach using combinations of Density Functional Theory, frozen phonon calculations, and phonon Boltzmann Transport theory to arrive at our final $\kappa$ values for the four monolayer honeycomb compounds of boron nitride, boron phosphide, boron arsenide, and boron antimonide. In Section II, we detail these computational methods. In Section III, our results and discussion are displayed. Finally, in Section IV we conclude our project and discuss future research.
\bigskip
\section{COMPUTATIONAL DETAILS}
\bigskip

All first-principles calculations are performed based on the Density Functional Theory (DFT) implemented using the ABINIT\cite{gonze2009} code. We utilize the Generalized Gradient Approximation (GGA) in the Perdew-Burke-Ernzerhof (PBE)\cite{perdew1996} parametrization as the exchange-correlation functional. We also utilize the pseudopotentials based on the Projected Augmented Wave (PAW) method.\cite{holzwarth2001} A self-consistent total energy tolerance criteria of 1.0$^{-10}$ Ha is used. The energy cutoffs used for the expansion of plane-wave basis sets, Monkhorst-Pack\cite{monkhorst1976} k-point grids, and vacuums are considered to be converged when total energy differences are less than $10^{-4}$ Ha twice, and are listed in Table \ref{structural}. Structures are fully relaxed when Hellman-Feynman forces are less than 1.0$^{-5}$ Ha/Bohr.

\begin{table}[ht]
\centering
\begin{adjustbox}{width=.5\textwidth}
\begin{tabular}{ | l | c | c | c | c  | c |}
\hline
Material & Energy Cutoff (Ha) & k-mesh & Vacuum (Bohr) & \textit{a} (Bohr)\\
\hline
BN& 25 & 6 $\times$ 6 $\times$ 1 & 14 &  4.75 (4.75\cite{sahin2009})\\
\hline
BP& 20 & 4 $\times$ 4 $\times$ 1 & 15 & 6.07 (6.07\cite{xie2016})\\
\hline
BAs& 13 &8 $\times$ 8 $\times$ 1& 16 &6.40 (6.41\cite{xie2016})\\
\hline
BSb& 14& 10 $\times$ 10 $\times$ 1 &9&7.05 (7.07\cite{xie2016})\\
\hline
\end{tabular}
\end{adjustbox}
\caption{Obtained kinetic energy cutoffs on the plane-wave basis, Monkhorst-Pack k-meshes, vacuums, and lattice constants as compared with other theoretical studies.}
\label{structural}
\end{table}

We employ the finite-displacement\cite{esfarjani2008} -- or frozen phonon approach -- to generate supercells:

\begin{equation}
\Phi_{\alpha\beta}(jl, j'l') \simeq -\frac{
F_\beta(j'l';\Delta r_\alpha{(jl)}) - F_\beta(j'l')} {\Delta
r_\alpha(jl)},
 \label{phonopy}
\end{equation}

In order to calculate forces, supercells are generated using Eq. \ref{phonopy}, where $\alpha$ and $\beta$ represent Cartesian indices, \textit{j} and \textit{j}' represent indices of atoms in the unit cell (before and after displacement, respectively), \textit{l} and \textit{l}' are indices of the unit cell (before and after displacement, respectively), $\Delta r_\alpha{(jl)}$ is the finite displacement, and $F_\beta(j'l';\Delta r_\alpha{(jl)}$ are the forces on atoms. 4 $\times$ 4 $\times$ 1 and 3 $\times$ 3 $\times$ 1 supercell sizes are chosen for second and third order displacements, respectively. For both sets of calculations, a 4 $\times$ 4 $\times$ 1 k-mesh is used. Phonon dispersion and group velocities are obtained from harmonic second order supercells. 

The intrinsic lattice thermal conductivity is calculated within the Phono3py\cite{phono3py}package based on anharmonic force constants inputted into the Boltzmann Transport Equation (BTE) for phonons. We employ the single-mode relaxation time approximation (RTA) to solve the BTE and to obtain the $\kappa$ tensor,

\begin{equation}
\kappa = \frac{1}{NV_0} \sum\limits_{\lambda}C_{\lambda}\textbf{v}_{\lambda}\otimes\textbf{v}_{\lambda}\tau_{\lambda}^{RTA}
\end{equation}

\noindent where $V_0$ is the unit cell volume, $\lambda$ describes phonon mode, $C_{\lambda}$ is the mode-dependent specific heat capacity, \textbf{v}$_{\lambda}$ is the modal group velocity, and $\tau_{\lambda}$ is phonon relaxation time at phonon mode $\lambda$.\cite{srivastava1990}

The thermal conductivity is calculated and normalized by multiplying results by $L_z/d$, where $L_z$ is the unit cell length along the z-direction, and $d$ is the thickness between layers, which is taken as the van der Waal's diameter of the atoms.\cite{qin2017} This method of normalization is well accepted for RTA calculations with two-dimensional materials.\cite{wu2017}

\bigskip
\section{RESULTS AND DISCUSSION}
\bigskip
	In this section, we report our results on various different phonon and thermal transport properties and analyze the underlying causes of their behavior. We first analyze the lattice dynamics properties, such as phonon dispersions, Debye temperature, and group velocities. These quantities reflect the results obtained from analysis of harmonic interatomic force constants. Then, we delve deeper into comparing our $\kappa$ results and offer many comparisons from other materials.
	
\begin{figure}[ht]
\centering
\begin{minipage}{0.3\textwidth}
	\subcaption{}
	\includegraphics[width=1\linewidth]	{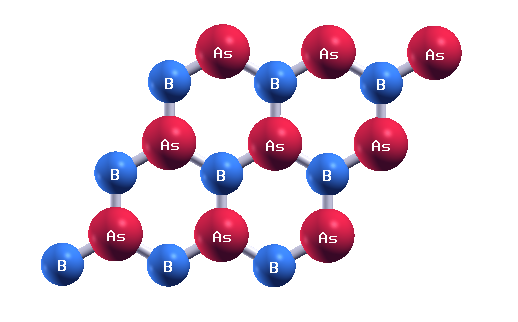}	
	\end{minipage}
   \begin{minipage}{0.3\textwidth}
 	\subcaption{}
	 \includegraphics[width=1\linewidth]{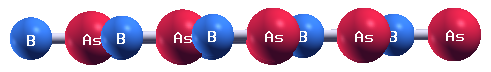}
   \end{minipage}
   \caption{(a) Top view (b) Side view of Monolayer BAs, which has the same planar crystal structure as monolayer BP and monolayer BSb.}
   \label{atomstruct} 
\end{figure}

\subsection{Lattice Dynamics}

\begin{figure*}[t]
    \centering
    \begin{subfigure}{0.5\textwidth}
\caption{}      
      \centering
      \includegraphics[width=\linewidth]{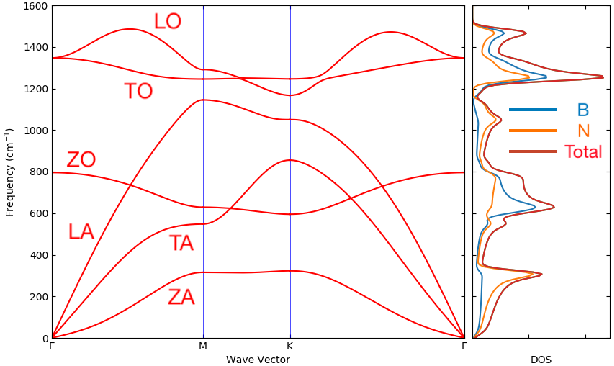}
    \end{subfigure}%
    \begin{subfigure}{0.5\textwidth}
\caption{}      
      \centering
      \includegraphics[width=\linewidth]{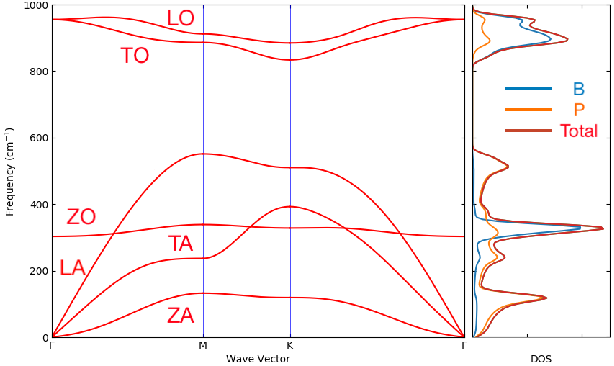}
    \end{subfigure}
    \begin{subfigure}{0.5\textwidth}
\caption{}
      \centering
      \includegraphics[width=\linewidth]{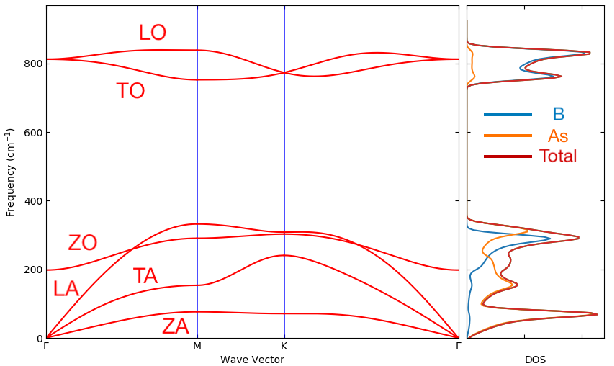}
      
    \end{subfigure}\begin{subfigure}{0.5\textwidth}
      \caption{}
      \centering
      \includegraphics[width=\linewidth]{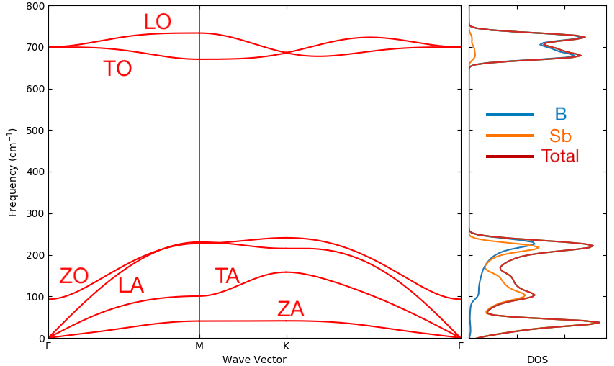}

    \end{subfigure}
  \caption {Phonon dispersions of monolayer hexagonal (a) BN (b) BP (c) BAs (d) BSb}
  \label{phononband}
\end{figure*}

\bigskip
\subsubsection{Crystal Structure and Phonon Dispersion}
\bigskip

All materials tested in our study follow the same planar honeycomb structure as seen in Fig. \ref{atomstruct}a, which shows the relaxed atomic structure of the monolayer h-BAs compound, corresponding to the \textit{P$\overline{6}$m2} space group. The space group is smaller compared to graphene because of the different atoms in the supercell. From Fig. \ref{atomstruct}b, it is evident that the structure is planar like graphene, with no buckling geometry as seen in structures such as stanene\cite{peng2016low} or phosphorene.\cite{jain2015}

Fig. \ref{phononband} shows the phonon dispersions of h-BAs and the other boron-V compounds. The two-atom unit cell corresponds to six phonon branches, due to each atom's three degrees of freedom for vibration. Since there has not yet been any experimental synthesis to our knowledge of monolayer BP, BAs, and BSb compounds, it is firstly very important to predict their dynamical stability based on phonon dispersion graphs. Since there exists no imaginary frequencies appearing in any of the phonon modes, we determine all three compounds to be structurally stable. In addition, they agree well compared to other \textit{ab initio} studies.\cite{sahin2009, xie2016}

However, in our case it is important to look at these phonon dispersions beyond their implications of stability. It is common knowledge that acoustic phonons are the main carriers of heat in phononic thermal transport, and thus analysis of those branches are of most importance. From all four compounds, we observe low frequencies of the longitudinal acoustic (LA), transverse acoustic (TA), and flexural acoustic (ZA) branches compared to the longitudinal optical (LO) and transverse optical (TO) branches. The out-of-plane optical (ZO) phonon modes have especially low frequencies in all of the compounds, and this pattern is observed in other planar two dimensional materials like graphene, h-AlN, h-GaN, and h-InN.\cite{sahin2009} 

We also observe a notably large frequency gap ($\omega_{gap}$) between the acoustic and some optical branches in h-BAs, which eliminates the resistive scattering channel where two acoustic phonons scatter to form one optical phonon (a + a $\rightarrow$ o). The more this channel is restricted by the frequency gap, the greater the thermal conductivity.\cite{broido2013} This $\omega_{gap}$ is due to the mass imbalances of the elements, as boron is a relatively atomically light element (10.81 amu), but arsenic (74.922 amu) is considerably heavy. We observe that the  $\omega_{gap}$ increases monotonically in all four compounds with average atomic mass, as displayed in Table \ref{harmonicprop}. The remarkable $\kappa$ in bulk BAs is largely attributed to its $\omega_{gap}$.\cite{broido2013} Since this large $\omega_{gap}$ is mirrored in the monolayer form, the high thermal conductivity we observe must also have its roots in the gap. It is also considerably important to note that h-BAs offers the ideal balance between large average atomic mass and large frequency gap, as h-BSb has a larger $\omega_{gap}$, but does not compensate enough for the larger average atomic mass.

Analysis of the partial density of states (PDOS) is also important to understand contributions to thermal conductivity. In all four dispersions, it can be seen that the boron atom contributes the most to optical modes, and the group V atoms contribute the most to acoustic modes. Consequently, the group V atoms thus have more contribution to thermal transport and their atomic masses play a large role in restricting thermal conductivity.

\begin{table}[ht] 
\begin{adjustbox}{width=.5\textwidth}
\begin{tabular}{| l |c|  c|  c|  c|  c | c| }
\hline
SL Structure & $\overline{M}$  (a.u.) & $\omega_{gap}$ (cm$^{-1}$)& $\Theta_D$ (K) & $v_{TA}$ (m/s) & $v_{LA}$ (m/s) \\
\hline
BN & 12.41 &100 &1230.34 & 10,819.62 & 20,079.30\\
\hline
BP&20.89&350&564.87& 6,895.79 & 13,267.33\\
\hline
BAs& 42.87 & 420 &435.62 & 4,972.66 & 9,234.13\\
\hline
BSb& 66.29& 440 & 329.19 & 3,691.05 & 7,099.68\\
\hline 
\end{tabular}
\end{adjustbox}
\noindent \caption{Average atomic masses ($\overline{M}$), frequency gaps ($\omega_{gap}$), Debye temperatures ($\Theta_D$), sound velocity of the TA phonon mode ($v_{TA}$), and sound velocity of the LA phonon mode ($v_{LA}$)}
\label{harmonicprop}
\end{table}

\bigskip
\subsubsection{Debye Temperature}
\bigskip

In Table \ref{harmonicprop} we compare some important features of each of the materials tested. Average atomic mass is displayed in order to draw distinctions and analyze its potential impact on thermal conductivity as theorized by Slack in 1973. Since h-BAs has a higher thermal conductivity than h-BN despite its larger average atomic mass, we prove that this is a contradiction of traditional Slack's theory, which agrees well with trends of the bulk form.\cite{lindsay2013}

The Debye temperature offers quantitative understanding on the harmonic  thermal conductivity, and can be approximated with the Debye model:

\begin{equation}
\Theta_D=\frac{\hbar \omega_{D}}{k_B}
\end{equation}

\noindent where $\hbar$ is Planck's constant, $\omega_{D}$ is the Debye frequency (the highest frequency of normal mode vibration), and $k_B$ is Boltzmann's constant.\cite{ashcroft1976} Following Slack's theory, a high Debye temperature correlates with a high thermal conductivity because it quantifies low average atomic mass and strong interatomic bonding.\cite{slack1973} In addition, the Debye temperature also represents the temperature at which all phonon modes begin excitation and below which phonon modes are frozen out. This is especially important when considering phonon-phonon collisions, where resistive Umklapp scattering takes over at temperatures higher than the Debye temperature. 

Debye temperatures ($\Theta_D$) of the materials are also listed in Table \ref{harmonicprop}. We find that $\Theta_D$ decreases monotonically from h-BN to h-BP to h-BAs to h-BSb with increasing average atomic mass, which is consistent with traditional law. The thermal conductivities also decrease from h-BN to h-BP to h-BSb, but h-BAs is the anomaly. Analogous to bulk c-BAs, monolayer h-BAs has a surprisingly low Debye temperature of 435.62 K when compared to h-BP (564.87 K) and h-BN (1230.34 K) despite its higher thermal conductivity. However, its frequency gap exceeds that of h-BN by 320 cm$^{-1}$ and h-BP by 70 cm$^{-1}$ as discussed earlier. This further proves the significance of the frequency gap as a crucial aspect in high thermal conductivity materials despite a low $\Theta_D$.
\bigskip
\subsubsection{Group Velocities}
\bigskip
\begin{figure*}[ht]
    \centering
    \begin{subfigure}{0.5\textwidth}
      \centering
      \includegraphics[width=\linewidth]{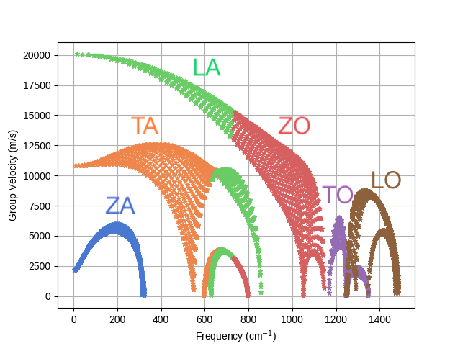}
      \caption{}
    \end{subfigure}%
    \begin{subfigure}{0.5\textwidth}
      \centering
      \includegraphics[width=\linewidth]{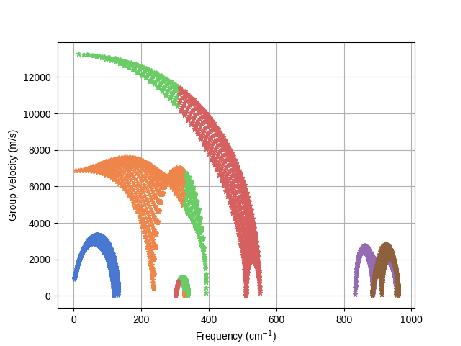}
      \caption{}
    \end{subfigure}
    \begin{subfigure}{0.5\textwidth}
      \centering
      \includegraphics[width=\linewidth]{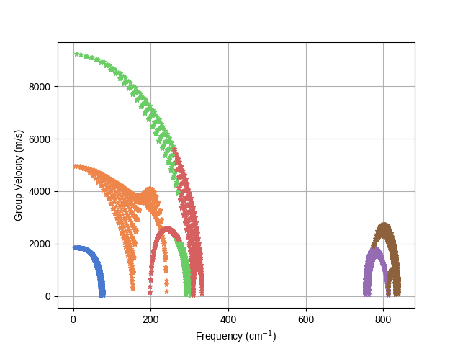}
      \caption{}
    \end{subfigure}\begin{subfigure}{0.5\textwidth}
      \centering
      \includegraphics[width=\linewidth]{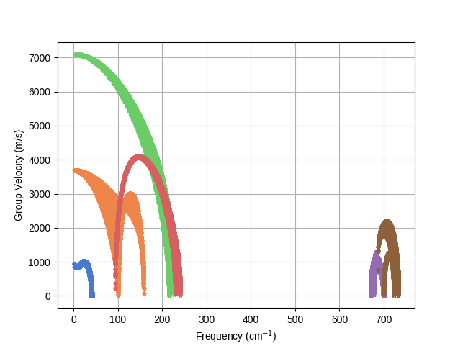}
      \caption{}
    \end{subfigure}
  \caption {Group velocities within the first Brillouin zone of (a) BN (b) BP (c) BAs (d) BSb }
  \label{group}
\end{figure*}

Phonon group velocities play an important role in determining the final thermal conductivity of materials. Not only do they offer insight on how much each phonon mode contributions to $\kappa$, but correlates directly to $\kappa$ by kinetic theory analysis:
\begin{equation}
\kappa = \frac{1}{3}{C_v}\textit{c}\textit{l}
\label{kinetic}
\end{equation}
where ${C_v}$ is specific heat, \textit{c} is phonon group velocity, and \textit{l} is the phonon mean free path.\cite{ashcroft1976} One can relate group velocity as the slopes at particular band points on the phonon dispersions. Thus when there is an evident frequency gap in the three materials as observed in Fig. \ref{phononband} (b-d), the same frequency gap is reflected in Fig. \ref{group} (b-d). This relation can simply be modeled as

\begin{equation}
v_g \equiv \frac{\partial \omega}{\partial \textbf{q}}
\end{equation}

\noindent where $\omega$ is the phonon mode frequency and \textbf{q} is the wave vector. Thus we can infer that larger frequency ranges within acoustic phonon branches in the dispersions will correspond to steeper slopes, and, consequently higher modal group velocities. The slope of acoustic branches near the $\Gamma$ wave vector signifies the material's sound velocity. From the graphs, the sound velocities are determined and displayed in Table \ref{harmonicprop}.

In h-BAs we see average group velocities to be much less than those of h-BN and h-BP. This is reflected in the low Debye temperature, and is caused by the smaller slopes due to smaller frequency ranges. The sound velocities of the LA and TA modes in h-BAs are nearly two times smaller in magnitude compared to h-BN. Resembling Debye temperature, the sound velocities also decrease monotonically from h-BN to h-BP to h-BAs to h-BSb. Traditionally, lower average group velocities imply a lower thermal conductivity, as relevant in Eq. \label{kinetic}, but we once again observe that this is not the case in h-BAs. This contradiction is also likely due the large frequency gap, which compensates for lower group velocities.

\bigskip
\subsection{Thermal Conductivity}
\smallskip

\begin{figure}[hpt]
\centering
\includegraphics[width=0.5\textwidth]{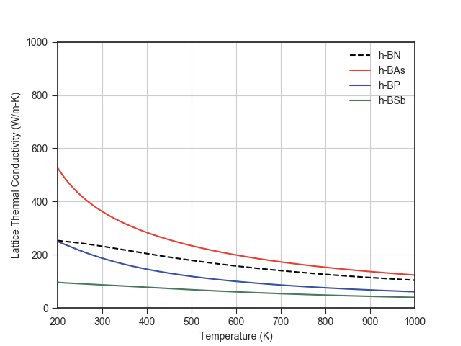}
\caption{Temperature dependence of lattice thermal conductivity in boron-V compounds}
\label{kappa}
\end{figure}

After understanding the impact of phonon dispersions and group velocities, we look deeper into our final results. Fig. \ref{kappa} shows the comparison of thermal conductivity in all materials tested versus temperature. Per traditional laws, the thermal conductivity decreases with increasing temperature due to resistive Umklapp processes being more active. We also observe the order of $\kappa$ from highest to lowest, namely h-BAs, h-BN, h-BP, and h-BSb. This order is identical to that of bulk structures.\cite{lindsay2013} As mentioned earlier, h-BAs has a large frequency gap which overcompensates for its large average atomic mass, enabling it to exceed the h-BN values. 
\bigskip
\subsubsection{Validity of our Thermal Conductivity Model}
\bigskip
It is widely known that the single mode relaxation time approximation (RTA) solution to the Boltzmann Transport Equation considerably underpredicts materials with high $\kappa$, as relevant in graphene\cite{d2017}. This is due to the inability of the RTA to detect differences between Umklapp and boundary phonon scattering.\cite{gu2014} Umklapp scattering, or U-processes, is the main phonon scattering mechanism limiting $\kappa$ in nonmetallic crystals. The RTA assumes U-processes to dominate at all temperatures, however this is not the case at low temperatures where nonresistive normal boundary scattering (N-processes) takes place. Thus the RTA underpredicts b   as a result of inadequate differentiation between N and U processes.

The calculated thermal conductivity of our basis material of monolayer h-BN differs drastically from the inaugural study that reported $\kappa$ to be around 600 W/m-K.\cite{lindsay2011} A more recent variational solution to the BTE in 2015 has shown $\kappa$ of h-BN to exceed even 1,000 W/m-K,\cite{cepellotti2015} whereas an iterative solution to the BTE in 2017 predicted the $\kappa$ to be around 245 W/m-K.\cite{qin2017} This study's h-BN $\kappa$ agrees well with Li et. al's study (both 232 W/m-K) in 2017,\cite{li2017} as we both employ the RTA for solving the BTE. In context of experimental studies, our calculated $\kappa$ most closely resembles five-layer h-BN, with a $\kappa$ of 250 W/m-K at room temperature. Moreover, bilayer hexagonal boron nitride was experimentally shown to have a $\kappa$ of 484 W/m-K. Despite these, there has not been any experimental studies on thermal conductivity for monolayer h-BN, and thus the $\kappa$ of h-BN is still in dispute. Our model can be considered accurate because it matches recent studies completed within the last two years.

\bigskip
\subsubsection{Context of Other Materials}
\bigskip

Many of Sahin et. al's massive collection of dynamically stable honeycomb compounds in 2009 have been probed for their thermal conductivity.\cite{sahin2009} Besides the commonly examined materials of graphene, silicene, germanene, and h-BN, intriguing allotropes of bulk covalent compounds have also been tested. This includes, but is not limited to, monolayer GaN, GeC, and AlN. The $\kappa$ of h-BAs scores high when compared to these materials. One of the closest contenders, monolayer AlN, has a predicted thermal conductivity of 264.10 W/m-K at room temperature.\cite{zhao2016} Although this is higher than h-BN likely due to its frequency gap, h-BAs offers a better combination of frequency gap and average atomic mass to overcome h-AlN in thermal conductivity. 

An example where frequency gap does not compensate for high average atomic mass is monolayer GaN,\cite{qin2017} where the heavier group III atom and smaller frequency gap compared to h-BAs is detrimental to the thermal conductivity. This is the same case for h-BSb, where the average atomic mass surmounts the frequency gap. Thus monolayer BAs stands at the perfect middle ground between too small of a frequency gap (h-BP) and too large of an average atomic mass with suboptimal compensation from frequency gap (h-GaN, h-BSb)

\bigskip
\subsection{Cumulative Thermal Conductivity}
\smallskip
\begin{figure}[ht] 
\centering
\includegraphics[width=0.5\textwidth]{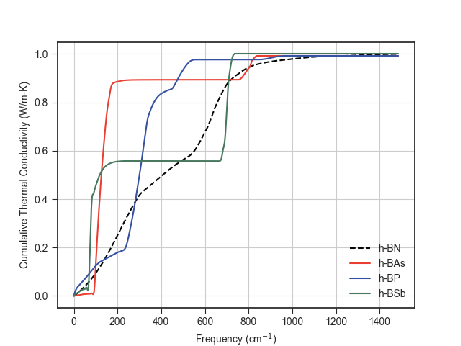}
\caption{Cumulative thermal conductivity of h-BP and h-BAs}
\label{kaccum}
\end{figure}

Fig. \ref{kaccum} shows the cumulative thermal conductivity with respect to phonon frequency. It is evident that low acoustic frequencies contribute more to thermal conductivity, and that the frequency gap is attributed to no contribution to the the thermal conductivity. Because of the lack of a substantial frequency gap in h-BN, the $\kappa$ steadily accumulates in the acoustic frequency ranges. However, in both h-BAs and h-BSb, accumulation spikes at low frequencies due to low acoustic frequencies. It is also interesting to see a spike in the 600-800 cm${-1}$ frequency range for h-BSb, as this range is occupied by optical phonon modes. Thus a factor limiting the thermal conductivity in h-BSb was likely its high optical phonon mode contribution.

\bigskip
\section{CONCLUSION} 
\bigskip

In summary, we investigated the two-dimensional forms of boron-V compounds for thermal conductivity and found that monolayer boron arsenide has higher thermal conductivity than its other boron-V cousins. Because of its electronic band gap and high $\kappa$, integration of monolayer boron arsenide into next-generation technology would prove to be possible and impactful. Monolayer boron-V compounds such as boron phosphide, arsenide, and antimonide are infrequently tested today both experimentally and theoretically. A deeper look into their thermal and mechanical properties would likely reveal captivating trends such as the ones we observed in this study. For example, in boron nitride it was shown that equibiaxial strains significantly enhanced thermal conductivity.\cite{li2017} Since the materials we tested today are of the same class as boron nitride, equibiaxial strains might enhance thermal conductivity as well. When integrating materials into electronic devices, it is crucial to understand how they conduct heat and its effect on device lifetime. The discovery that boron arsenide has relatively high thermal conductivity for monolayer structures will likely have unprecedented applications in thermal management and electronics cooling.

\bigskip
\section{ACKNOWLEDGEMENTS}
\bigskip
We would like to thank Dr. Gefei Qian, who helped us with technical requirements.
We would also like to thank Dr. Wu Li, Dr. Suhuai Wei, Dr. Jihui Yang, Ms. Shasha Li, and Dr. Hiroshi Uchiyama for their very fruitful discussions.

\newpage
\bibliography{sample_ref}{}
\end{document}